\documentclass[np2,twoside,twocolumn,showpacs]{revtex4}

\usepackage{amssymb}
\usepackage{amsmath}
\usepackage{graphicx}
\usepackage{epsfig}
\usepackage{bm,multirow}

\def\NESTED{\mbox{\begin{picture}(7,11)
\put(1,10){\line(1,0){10}}
\put(1,0){\line(1,1){10}}
\put(1,0){\line(0,1){10}}
\end{picture}
}}

\def\CPSHAPE{\mbox{\begin{picture}(7,11)
\put(1,10){\line(1,0){10}}
\put(1,0){\line(0,1){10}}
\put(1,0){\line(1,0){5}}
\put(6,0){\line(0,1){5}}
\put(6,5){\line(1,0){5}}
\put(11,5){\line(0,1){5}}
\end{picture}
}}

 \volumenumber{xx} \issuenumber{x}
\volumeyear{Month Year} 
\startpage{1}
\endpage{5}

\begin{document}

\title[]{Mesoscale Properties of Mutualistic Networks in Ecosystems}

\author{Sang Hoon \surname{Lee}} 
\affiliation{Department of Physics and Research Institute of Natural Science, Gyeongsang National University, Jinju 52828, Korea}
\affiliation{Future Convergence Technology Research Institute, Gyeongsang National University, Jinju 52849, Korea}

\date[]{Submitted \today}

\begin{abstract}
Uncovering structural properties of ecological networks is a crucial starting point of studying the system's stability in response to various types of perturbations. We analyze pollination and seed disposal networks, which are representative examples of mutualistic networks in ecosystems, in various scales. In particular, we examine mesoscale properties such as the nested structure, the core-periphery structure, and the community structure by statistically investigating their interrelationships with real network data. As a result of community detection in different scales, we find the absence of meaningful hierarchy between networks, and the negative correlation between the modularity and the two other structures (nestedness and core-periphery-ness), which themselves are highly positively correlated. In addition, no characteristic scale of communities is perceivable from the community-inconsistency analysis. Therefore, the community structures, which are most widely studied mesoscale structures of networks, are not in fact adequate to characterize the mutualistic networks of this scale in ecosystems. 
\end{abstract}


\maketitle

\title[]{생태계 상호 도움 네트워크의 중간 크기 성질 분석}

\author{이상훈}
\email{lshlj82@gnu.ac.kr} 
\affiliation{경상국립대학교 물리학과 및 기초과학연구소, 진주 52828, 대한민국}
\affiliation{경상국립대학교 미래융복합기술연구소, 진주 52849, 대한민국}

\date[]{\today~투고}

\begin{abstract}
생태계 네트워크의 구조적 성질을 밝혀내는 것은 외부 환경 변화, 종의 개체수 등락과 같은 내부 환경 변화에 대한 계의 안정성을 추론할 수 있는 중요한 출발점이다. 본 연구에서는 생태계에 존재하는 대표적인 상호 도움 네트워크인 곤충과 새에 의한 식물의 수분과 종자 분산 네트워크를 다양한 척도에서 분석하였다. 특히, 생태계 복잡성이 발현되는 지점이라고 할 수 있는 네트워크의 중간 크기 성질들로 제시되어 온 포개진 구조, 중심-주변부 구조, 군집 구조들이 서로 어떻게 연관되어 있는지를 실제 네트워크 자료를 이용하여 통계적으로 살펴보았다. 다양한 척도에서 나타나는 군집 구조를 이용한 네트워크들 간의 분류 방법을 통한 계층 구조 추론을 한 결과 수분 네트워크와 종자 분산 네트워크가 구분이 되지 않음을 발견하였다. 이전 연구에서 서로 깊은 상관관계가 있다고 밝힌 바 있는 포개진 구조와 중심-주변부 구조 같은 것들이 군집 구조와는 음의 상관관계가 있음을 보였고, 작은 척도에서 즉 세부적인 군집 구조에서 음의 상관관계가 더 깊어짐을 발견했다. 또한 군집 구조들의 불일치 정도를 통한 안정성 분석에서도 특징적인 척도를 발견하지 못하였다. 즉, 이 정도 크기의 생태계 상호 도움 네트워크들을 기술하는 데 있어서는 가장 흔히 분석되는 중간 크기 성질인 군집 구조가 적합하지 않은 것으로 보인다.
\end{abstract}


\maketitlekorean
 
  

\pagebreak

\setcounter{footnote}{0}

\section{서론}
\label{sec:introduction}

생태계의 상호작용\cite{Roberts1974,Bastolla2009,Allesina2012,Rohr2014}은 자연계에 존재하는 대표적인 연결망 또는 네트워크 구조\cite{NetworkReview}로서 포식자-피식자 관계 등의 적대적 관계, 공생 관계 등의 긍정적인 상호 의존적 관계를 모두 포함한다. 네트워크 과학\cite{NetworkReview}은 상호작용 자체에 대한 연구 분야로서 그러한 상호작용을 연구하기에 적합한 분석 방식들을 제공하며, 특히 네트워크에서의 동역학 과정\cite{Porter2016}이나 고윳값 (eigenvalue) 스펙트럼 분석\cite{Allesina2012,May1972,Majumdar2014} 연구는 생태계에서 매우 중요한 종의 개체수 변화와 안정성을 예측하는 데 필수적이다. 특히 그 중에서도 상호작용의 주체들의 연결 관계가 일반적인 것들 (generalists)이 세부적인 것들 (specialists)의 연결 관계를 계층적으로 포함한다는 뜻의 포개짐 (nestedness)\cite{NestednessReview}은 생태계나 경제 네트워크를 특징짓는 중요한 중간 크기 (mesoscale) 성질로 널리 알려져 있다. 다만, 최근에 이러한 포개진 구조가 고윳값 스펙트럼 분석 결과로는 생태계 안정성에 오히려 좋지 않을 수 있다는 연구 결과들\cite{Allesina2012,Staniczenko2013}이 발표되고 있고, 본 저자도 참여한 연구들에서 이러한 포개진 구조가 독립적인 구조라기보다는 연결선수 (degree) 분포에 의해 결정되거나\cite{SHLee2016,Bruno2020} 더 근본적으로 또다른 중간 크기 성질인 중심-주변부 (core-periphery) 구조\cite{Csermely2013,Rombach2014,SHLee2014,Cucuringu2016}로부터 파생되었을 가능성이 제시된 바 있다\cite{MartinGonzalez2020}. 즉 생태계 네트워크에서 이러한 중간 크기 성질의 정확한 역할에 대해서는 아직 밝혀지지 않은 것이 많은 것이다.

이러한 상황에서 실제로 생태계 네트워크의 구조적 성질 자체에서도 과연 포개진 구조나 중심-주변부 구조 외에 다른 중간 크기 성질은 없는지 충분히 점검할 필요가 있고, 그 중에서 특히 가장 널리 알려진 분석 틀인 군집 구조 (community structure)\cite{CommunityReviewPorter,CommunityReviewFortunato}를 앞서 언급한 다른 중간 크기 구조적 성질과 연관지어 분석해 보는 것이 본 연구의 목적이다. 본 저자는 이전 연구\cite{SHLee2016}를 통해 ``web of life'' 데이터베이스에서 제공하는 89개의 생태계 상호 도움 네트워크\cite{WebOfLife}에서의 포개진 구조와 중심-주변부 구조 사이의 상관관계를 밝힌 바 있는데, 본 연구에서는 그 네트워크들을 다양한 척도에서 나타나는 군집 구조 관점에서 살펴봄으로써 다른 측면에서의 구조적 특성이 있는지를 살펴보았다. 군집 구조 자체와 포개진 구조, 중심-주변부 구조와의 상관관계, 또한 군집 구조로부터 파생된 다양한 분석법, 예를 들어 네트워크들 간의 계통수\cite{Onnela2012} 분석이라든지 검출된 군집들 사이의 불일치도 (inconsistency) 분석\cite{HKim2019,DLee2020}을 통한 분석 결과 특정한 척도를 가진 의미 있는 군집 구조를 찾기 힘들었으며, 따라서 포개진 구조, 중심-주변부 구조와는 달리 적어도 이 정도 크기를 단위로 하는 생태계의 공생 관계를 나타내는 도움 네트워크에서는 군집 구조를 찾는 것이 큰 의미가 없다는 결론을 도출하였다.

\section{연구 방법 및 데이터}
\label{sec:method_and_data}

\subsection{포개진 구조와 중심-주변부 구조}
\label{sec:nestedness_and_c_p}

생태계 네트워크의 구조적 성질로서 총설논문\cite{NestednessReview}도 최근에 발표될 정도로 많이 연구된 것이 포개진 구조 (nested structure)이다. 포개진 구조는 네트워크를 연결선수 순서대로 내림차순 정렬했을 때 연결선수가 상대적으로 많은 노드 $i$와 상대적으로 적은 노드 $j$가 있다고 했을 때, $j$와 상호작용하는 노드들이 대체로 $i$와도 상호작용을 한다는, 즉 노드 $j$와 연결된 이웃 노드 집합 $\nu(j)$가 노드 $i$의 이웃 노드 집합 $\nu(i)$의 진부분집합 (proper subset)인 경향이 있다는 것이다. 수학적으로 쓰자면, 흔히 사용하는 표기법에 따라 $k_i$를 노드 $i$의 연결선수라고 하면 $k_i = |\nu(i)| > k_j = |\nu(j)|$일 때 $\nu(j) \subset \nu(i)$가 대체로 성립한다고 표현할 수 있다. 이 때 서론 (\ref{sec:introduction} 절)에서 언급했듯이 연결선수가 많은 노드 $i$쪽이 연결 관계가 일반적인 것들 (generalists)에 해당되며, 연결선수가 적은 노드 $j$쪽이 연결 관계가 세부적인 것들 (specialists)에 해당한다. 시각적으로는 네트워크를 표현하는 인접행렬 (adjacency matrix) 성분 $W_{ij}$\footnote{가중치 없는 네트워크 (unweighted network)의 경우 노드 $i$와 노드 $j$의 연결이 있으면 $W_{ij} = 1$, 없으면 $W_{ij} = 0$이며, 가중치 (weight)가 있을 경우 가중치 값이 $W_{ij}$가 된다 (가중치가 있는 경우에도 연결이 없을 때는 보통 $W_{ij} = 0$으로 표현한다).}를 각 노드의 연결선수에 따라 행, 열 모두 내림차순으로 정렬했을 때 채워진 모양이 \NESTED~형태가 된다. 포개짐 정도를 정량화하는 측정량들 중 본 연구에서 사용한 것은 겹쳐짐과 내림차순 채움을 이용한 포개짐 정도 (nestedness metric based on overlap and decreasing fill: NODF)\cite{AlmeidaNeto2008,SHLee2016} $\nu$ 이며, 다음과 같이 정의된다.
\begin{equation}
\nu = \frac{\displaystyle \sum_{r=1}^{n(n-1)/2} f_{\mathrm{col}}(r) + \sum_{c=1}^{m(m-1)/2} f_{\mathrm{row}}(c)}{n(n-1)/2 + m(m-1)/2} \,.
\label{eq:NODF}
\end{equation}
여기에서 분자의 $f_{\mathrm{col}}(r) \in [0,1]$ 과 $f_{\mathrm{row}}(c) \in [0,1]$는 각각 인접행렬 성분 $W_{ij}$를 행과 열을 연결선수에 대한 내림차순으로 정렬한 후 행의 쌍 $r$에 대해 포개짐 구조를 만족하는 열의 비율, 열의 쌍 $c$에 대해 포개짐 구조를 만족하는 행의 비율을 나타내며 행의 수 $n$, 열의 수 $m$로부터 가능한 총 행과 열의 쌍 개수를 더한 것을 분모에 넣어 $\nu \in [0,1]$이 되도록 (포개짐 정도가 증가할수록 $\nu$가 증가하며 완벽하게 포개진 구조가 최댓값 $\nu = 1$이 되도록) 규격화한 것이다. 

생태계에서 이러한 구조가 많이 관찰된다\cite{NestednessReview}는 것이 알려지면서, 자연스럽게 이러한 구조가 생태계를 안정화시키기 위한 것이 아닌가 하는 추론이 있을 수 있고 관련 연구들이 많이 이루어졌다. 다만, 이 포개진 구조가 (존재한다고 가정하더라도) 과연 실제로 생태계의 안정성에 기여하는지\cite{Rohr2014}에 대해서는 여러 반론\cite{Allesina2012,Majumdar2014}이 제기되고 있다. 이것은 모형 연구에서 필연적으로 가정할 수 밖에 없는 단순화된 동역학계의 세부 구조에 따라 관점이 달라질 수 있는 부분이며, 다만 네트워크의 어떤 구조적 특성이 존재한다는 이유만으로 그 특성이 그 계를 안정화시키는 방향으로 기여한다고 생각해서는 안 된다고 본다. 예컨대 어떤 구조적 특성은 생태계 안정화보다는 다른 목적함수 (objective function)를 최적화하는 데 기여하고 있을 수도 있으며, 다른 구조적 특성이 그럼에도 불구하고 계를 안정화시키고 있을 수도 있다. 본 저자는 선행연구\cite{SHLee2016}를 통해 인접행렬 모양의 유사성 (\NESTED~ 와 \CPSHAPE~)에서 출발하여 포개진 구조와 중심-주변부 구조 사이의 깊은 상관관계를 보고한 바 있다. 여기서 더 나아가, 최근에는 각종 문헌 조사와 실제 분석을 통해 포개진 구조보다는 중심-주변부 구조가 더 근본적으로 생태계 네트워크를 기술하는 방식이 아닐까 하는 주장\cite{MartinGonzalez2020} 또한 제기한 바 있다. 

중심-주변부 구조\cite{Csermely2013,Rombach2014,SHLee2014,Cucuringu2016}는 네트워크의 노드들을 가장 중심 (core) 부분부터 가장 주변부 (periphery) 부분까지 적절하게 정렬\footnote{사실 많이 강조되지는 않지만 이 부분에서 일반적으로 군집 구조와 중심-주변부 구조를 살펴보는 관점의 차이가 있다. 대개 전자에서는 군집을 확실하게 나누는 것에 관심이 있는 반면, 후자에서는 중심과 주변부를 확실하게 나누는 것 보다는 좀 더 일반적으로 중심에서 주변부까지의 순서 또는 연속적으로 변하는 점수 (score)를 부여하는 경우가 많다. 물론 중심-주변부 구조에서도 적절한 문턱값과 목적함수를 주어서 최적화를 통해 중심과 주변부를 구분할 수도 있다\cite{Cucuringu2016}.}할 수 있으며 중심 부분끼리는 잘 연결되어 있고, 주변부 부분도 다른 주변부 부분이 아닌 중심 부분과 연결되려 하는 경향성을 뜻한다. 이러한 중심-주변부의 구분 정도를 나타내는 양으로 본 연구에서 사용한 중심-주변부도 (core quality: CQ) $\xi$는 다음과 같이 참고문헌\cite{SHLee2016}에서 정의한 바 있으며, 일반적인 네트워크 구조에서 정의된 양을 동물, 식물로 노드가 구분되며 연결은 언제나 동물과 식물 사이에만 있는 양자간 네트워크 (bipartite network)인 생태계 상호 도움 네트워크에 맞도록 수정한 것이다. 
\begin{equation}
\xi = \frac{\displaystyle \sum_{i,j} W_{ij} \Xi_\mathrm{animal} (i) \Xi_\mathrm{plant} (j)}{\displaystyle \sum_{i,j} W_{ij} \sum_{i,j} \Xi_\mathrm{animal} (i) \Xi_\mathrm{plant} (j)} \,.
\label{eq:NCQ_formula}
\end{equation}
여기에서 $W_{ij}$은 노드 $i$와 $j$의 연결을 나타내는 인접행렬이며, 동물 노드와 식물 노드에서 각각 정의되는 중심성 점수 (core score) $\Xi$는
\begin{equation}
\Xi_{\omega} (i) = Z_{\omega} \sum_{({\alpha},{\beta})} C_i ({\alpha},{\beta}) R({\alpha},{\beta}) 
\label{eq:CS_formula}
\end{equation}
로 각 노드 $i$에 대해 정의된다 ($\omega \in \{ \mathrm{animal}, \mathrm{plant} \}$). 여기에서 $\alpha \in [0,1]$, $\beta \in [0,1]$은 참고문헌\cite{Rombach2014}에서 정의한 바와 같이 중심-주변부의 전환이 얼마나 급격하게 이루어지는지와 중심 노드의 비율을 조절하는 매개변수들이며, 식에서 정의한 바와 같이 가능한 경우의 수에 대해 모두 가중치가 있는 평균을 낸 것이다. 본 연구에서는 이전 연구\cite{SHLee2016}와 같이, 동물과 식물 노드 모두 $\alpha = 1/2$로 고정시키고 $\beta$를 $0$에서 $1$사이에서 $\Delta \beta = 0.01$ 간격으로 변화시킨 값들을 이용하였다. 각 $\alpha$, $\beta$ 값에서 중심성 벡터 $C_i$는 노드 $i$가 얼마나 중심성을 가지고 있는지를 나타내는 두 선형함수로 이루어진 조각별 선형 (piecewise linear) 함수이며, $R$은 그 경우에 얼마나 중심-주변부의 기본 가정인 중심부들끼리 연결이 잘 되어 있는지를 정량화하는 양이다. (더 자세한 것은 \cite{Rombach2014,SHLee2016} 참고.) 중심-주변부 구조에서 중심 부분을 포개진 구조에서 일반적인 것들, 주변부 부분을 포개진 구조에서 세부적인 것들에 대응시키면 구조적으로 둘이 사실은 비슷한 구조를 기술하게 되며 결국 독립적인 이야기가 아니라는 것이 이전 연구\cite{SHLee2016}의 핵심 결과이다.

\subsection{군집 구조}
\label{sec:community_structure}

군집 구조 (community structure)\cite{CommunityReviewPorter,CommunityReviewFortunato}는 \ref{sec:nestedness_and_c_p} 절에서 소개한 포개진 구조, 중심-주변부 구조보다 훨씬 더 많이 연구된 네트워크의 중간 크기 성질 (mesoscale property)이다. 생태계 네트워크에서도 예외가 아니어서, 참고문헌\cite{Fortuna2010}과 같은 연구를 통해 생태계 네트워크의 포개진 구조가 군집 구조와 관련성이 있다는 주장이 있었다. 본 연구에서는 이전 연구들\cite{SHLee2016,MartinGonzalez2020}을 통해 밝힌 포개짐 구조 또는 중심-주변부 구조 외에도 과연 그러한 군집 구조가 생태계 상호 도움 네트워크에서 실제로 유의미한 것인지를 살펴보려고 한다.

군집 구조는 전체 네트워크를 몇 개 노드들의 집합으로 나누어지며 각 해당 집합에 속하는 노드들 간의 연결이 다른 집합에 속한 노드들과 비교해서 통계적으로 유의미하게 많은 것을 뜻한다. 이러한 뭉침 성질 (clustering property)을 찾는 것은 전체 계를 유의미한 크기의 작은 부분들로 쪼개서 연구하는 환원론적인 관점에서 출발한다고 볼 수도 있고, 각종 패턴을 (심지어 없는 경우에도) 찾는 것에 익숙한 인간의 사고방식에서 비롯된다고 볼 수도 있을 것이다\footnote{최근 기계학습 (machine learning) 분야에 있어서도 뭉침 성질을 이용한 분류는 매우 중요한 분야이다.}. 네트워크가 매우 작을 경우에는 인간이 본능적으로 잘 수행하는 일이지만, 네트워크의 크기가 조금만 커져도 전체 노드를 구분하는 일이 NP-난해 (NP-hard)\cite{CommunityReviewPorter,CommunityReviewFortunato} 문제가 되면서 수학적으로 잘 정의된 목적함수와 적절한 최적화 알고리즘을 써야 군집 구조를 찾을 수 있다. 

그러한 군집 구조를 찾을 때 가장 많이 이용되는 목적함수인 군집도(modularity)는 다음과 같이 정의된다\cite{CommunityReviewPorter,CommunityReviewFortunato}.
\begin{equation}
Q = \frac{1}{2m} \sum_{i \ne j} \left[ \left( A_{ij} - \gamma \frac{k_i k_j}{2m} \right) \delta(g_i,g_j) \right ] \,,
\label{eq:modularity}
\end{equation}
여기에서 $A_{ij}$는 인접행렬 (adjacency matrix)의 성분으로서 노드 $i$와 노드 $j$의 가중치 없는 네트워크 (unweighted network)의 경우에는 연결 여부(연결되었으면 $1$, 연결되지 않았으면 $0$)를 나타내고, 가중치 있는 네트워크 (weighted network)의 경우에는 $0$또는 $1$이 아닌 값을 허용하여 연결의 강도 (weight)를 나타낸다. 그 뒤에 나오는 영모형 (null model) 항인 $k_i k_j / (2m)$은 각 노드의 연결선수 (degree)의 곱을 연결선의 전체 개수인 $m$으로 나누어 (각 노드가 $i$와 $j$에 한 번씩 들어가므로 중복을 고려하여 $2$로 더 나눠준다) 적절하게 재규격화 (renormalize)한 것으로서, 군집 구조 (뿐만 아니라 연결선수 외에 어떤 구속조건도 없이 무작위로 연결되었다고 가정)가 없다고 가정했을 때 예상되는 노드 $i$와 노드 $j$의 연결확률이다. 가중치 있는 네트워크의 경우에도 $k_i = \sum_j A_{ij}$라는 같은 정의를 써서 일반화된 연결선수라고 할 수 있는 연결강도 (strength)를 나타내고 가중치의 총합을 $m$으로 하여 영모형 항의 의미를 같게 맞춰준다. 식~\eqref{eq:modularity}의 맨 뒤에 있는 크로네커 델타 (Kronecker delta) 함수에 있는 $g_i$는 노드 $i$가 속한 군집을 나타나며, 따라서 $\delta(g_i,g_j)$는 노드 $i$와 $j$가 같은 군집에 속하면 $1$이 되어 앞의 괄호 속 항이 군집도 계산에 포함되며, 다른 군집에 속하면 $0$이 되어 괄호 속 항이 군집도 계산에 포함되지 않는다. 

이 목적함수~\eqref{eq:modularity}를 잘 보면 네트워크가 정해지면 다른 항들은 모두 이미 결정되기 때문에 이 목적함수를 바꾸는 유일한 부분이 바로 그 군집을 어떻게 나눌 것인가에 해당되는 $\delta(g_i,g_j)$ 함수인 것이며, 따라서 이 때 군집도 $Q$를 최대화하는 군집 구조를 찾는 것이 네트워크에서 군집 구조 찾기 문제의 목적이 된다. 또한 이 목적함수~\eqref{eq:modularity}의 영모형 항 앞에 곱해진 $\gamma$는 어떤 크기의, 또는 얼마나 자세한 군집 구조를 찾을지를 결정하는 중요한 해상도 매개변수 (resolution parameter)의 역할을 한다\cite{Onnela2012}. 구체적으로 말하자면 해상도 매개변수 $\gamma$가 클수록 더 작은 크기의 군집들을 (군집의 개수 관점에서는 더 많은 수의 군집들을) 찾게 되는데, 본 연구에서 군집 구조를 찾을 때는 단순히 하나의 척도에서 군집을 찾는 게 아니라 여러 $\gamma$ 값으로부터 검출되는 군집 구조들을 살펴봄으로써 네트워크를 여러 척도의 중간 크기에서 조사할 것이다. 가장 처음으로 수행한 것은 다양한 척도에서 나타나는 군집의 특성을 이용하여 네트워크들 사이의 유사도 (또는 그것의 반대방향인 거리)를 정의하고 그것을 바탕으로 다양한 네트워크 집합을 계층적으로 구분하는 중간 크기 반응 함수 (mesoscopic response function: MRF) 분석\cite{Onnela2012}이다.

이렇게 목적함수를 하나로 결정하더라도 (당연히 실제로는 목적함수도 꼭 군집도 $Q$일 필요는 없다) 총설논문들\cite{CommunityReviewPorter,CommunityReviewFortunato}에 잘 나오듯이 어떤 최적화 과정을 통해 군집 구조를 찾을 것인가 하는 것은 열린 문제 (open question)이며, 그 무한한 가능성이 이 쪽 분야를 발전시키는 원동력 중 하나이기도 하다. 그러한 목적함수 최적화 방법은 크게 결정론적인 (deterministic) 방법과 확률적인 (stochastic) 방법으로 구분되는데 널리 사용되는 (그리고 본 연구에서도 사용할) Louvain\cite{Louvain} 알고리즘의 변형인 GenLouvain\cite{GenLouvain}과 같은 유명한 방법이 확률적인 방법에 해당한다. 확률적인 방법은 말 그대로 확률과정을 통해 결과가 나오기 때문에 매 시행마다 원칙적으로는 다른 결과가 나올 수 있고, 본 저자도 참여한 관련 연구들\cite{HKim2019,DLee2020}에서는 이 성질을 이용하여 군집 구조에 있어서의 각종 불일치도 (inconsistency)를 정량화하고 이를 이용하여 개별 노드, 군집 구조 자체의 안정성 등을 논의하였다. 본 연구에서도 살펴볼 다양한 척도 (이전 단락에서 언급한 여러 $\gamma$ 값) 에서의 군집 구조의 특성 중 하나로서 이 불일치도 성질을 이용할 예정이다.

\subsection{상호 도움 네트워크 데이터}
\label{sec:data}

분석에 이용할 대표적인 생태계의 상호 도움 네트워크 자료로, 기존 연구\cite{SHLee2016}에서도 분석한 바 있는 ``web of life''\cite{WebOfLife}의 $59$개의 수분 (pollination) 네트워크와 $30$개의 종자 분산 (seed dispersal) 네트워크로 구성된 총 $89$개의 네트워크를 이용하였다. 각 네트워크는 식물과 동물 (해당 식물의 수분 또는 종자 분산을 돕는 곤충 또는 새) 사이의 양자간 네트워크 (bipartite network)로 구성되어 있으며, 특히 중심-주변부 구조를 정량화하는 양은 \ref{sec:nestedness_and_c_p} 절에서 기술하였듯이 양자간 네트워크에 맞도록 변형하여 계산하였다.

\section{결과 분석}
\label{sec:result}

\subsection{군집 구조를 이용한 네트워크들 간의 계층 구조 파악}
\label{sec:MRF}

\begin{figure*}[t]
\includegraphics[width=\textwidth]{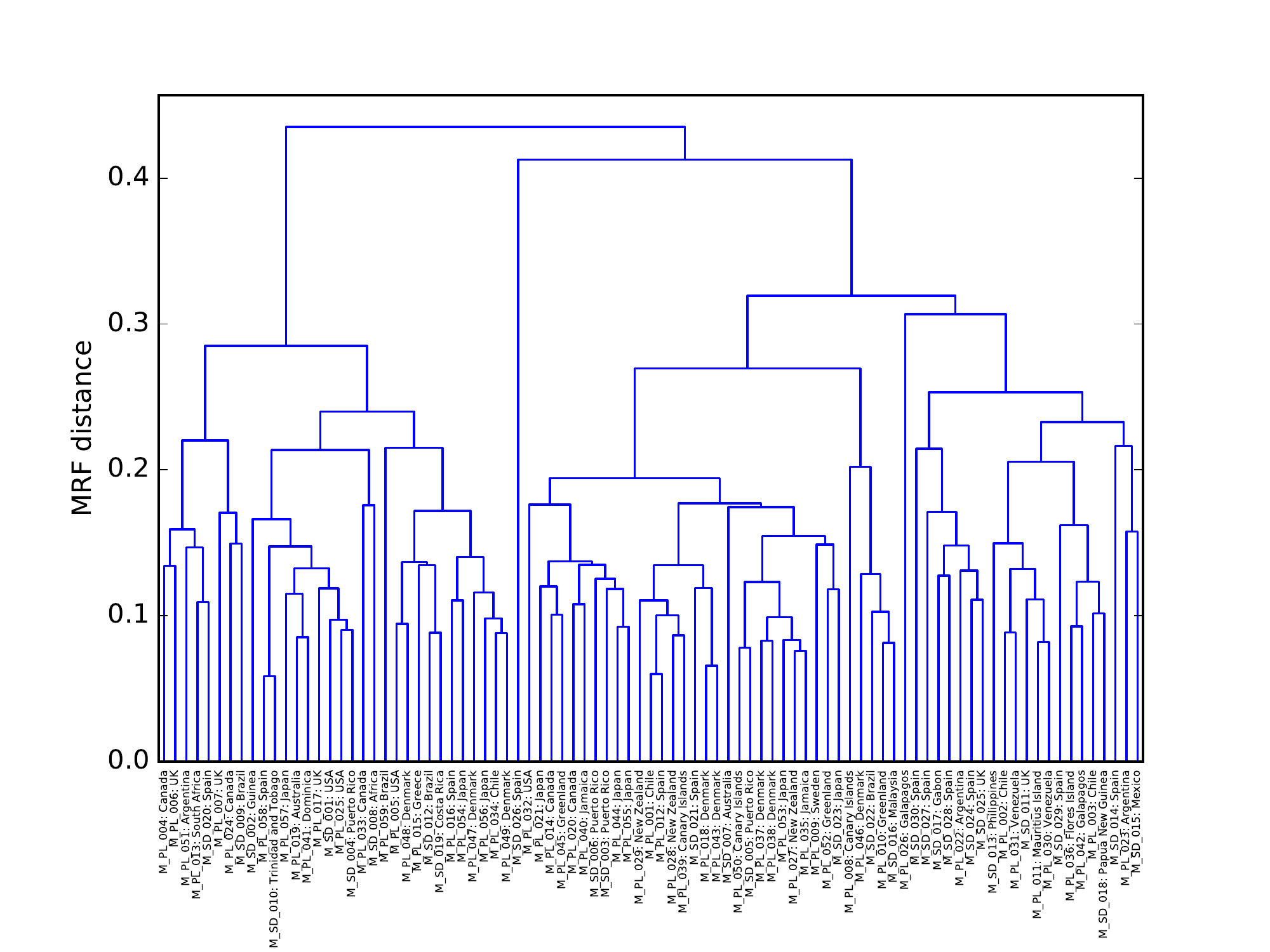}
\caption{The dendrogram based on the MRF distance~\cite{Onnela2012} for all of the $89$ mutualistic networks. The labels include either pollination (PL) or seed dispersal (SD) to classify the networks, followed by the countries or the regions where they are located.}
\label{fig:dendrogram}
\end{figure*}

네트워크의 군집 구조는 개별 네트워크 단위에서 하부 구조를 파악할 뿐만 아니라, 적절한 다시눈금잡기 (rescale) 과정을 통하여 다양한 척도에서 나타나는 군집들의 성질을 이용하여 다른 네트워크들 사이의 거리를 정의하고 그것을 바탕으로 네트워크들을 계층적으로 분류하는 데도 쓰일 수 있다\cite{Onnela2012}. 말하자면 현미경의 조동/미동 나사를 돌려가면서 초점을 맞추듯이 식~\eqref{eq:modularity}에 나오는 해상도 매개변수 $\gamma$를 조절하면서 크고 작은 척도에서 네트워크의 군집 구조를 찾아내 보고, 거기서 파생되는 각종 양들의 양상을 네트워크의 특성으로 기록할 수 있다. 참고문헌\cite{Onnela2012}에서는 이렇게 각 척도에 따라 발생하는 네트워크의 특성들을 3가지의 함수로 표현한 다음 두 개의 다른 네트워크에서 발생하는 이 3개의 중간 크기 반응 함수들 사이의 차이 절댓값을 해상도 변수로 적분한 다음 그 양을 그 두 네트워크 사이의 거리로 정의한다. 따라서 각 네트워크 쌍에 대해 3차원의 (비)유사도\footnote{덜 유사한 것들일수록 거리는 먼 것에 해당하기 때문에 이렇게 표현하였다.} 벡터를 계산할 수 있고, 주성분분석(principal component analysis)을 통하여 각 네트워크들 사이의 최종 거리를 계산하게 된다. 그리고 최종적으로 그 거리 쌍으로부터 계통수 (dendrogram)를 얻을 수 있다.

그림~\ref{fig:dendrogram}은 이러한 군집 구조 (군집을 찾는 방법으로는 참고문헌\cite{Onnela2012}에서 사용한 Potts 모형 방법\cite{Reichardt2006}을 이용하였다)를 이용한 $89$개 네트워크들 간의 계층 구조를 계통수로 나타낸 것이다. 계통수를 그리는 방식으로는 평균 결합 뭉치기 (average linkage clustering) 방법을 이용하였다. 이 $89$개 네트워크를 가장 크게 나누는 기준이 될 수분 네트워크와 종자 분산 네트워크도 계통수에서 거의 구분이 되지 않는 것을 볼 수 있다. 또한 비슷하게 묶인 네트워크들도 여러 대륙에 걸쳐 있는 것으로 보아 각 네트워크들이 위치한 지리적 위치와도 상관관계가 없음을 알 수 있다. 따라서 이것은 다양한 척도에서의 군집 구조를 이용하여 찾아낸 네트워크들의 계층 구조가 실제 외부 정보를 유의미하게 검출하지 못하고 있음을 뜻한다. 물론 이렇게 수분과 종자 분산에 따른 구분, 지리적 위치와는 다른 숨겨진 구조가 있을 수도 있지만, 이렇게 군집 구조가 찾아낼 수 있는 정보량이 부족한 것은 앞으로 제시할 다른 결과와도 일치하는 방향이다.

\subsection{다양한 척도에서의 군집 구조와 알려진 구조들 사이의 상관관계}
\label{sec:correlation}

\begin{figure}[t]
\includegraphics[width=0.5\textwidth]{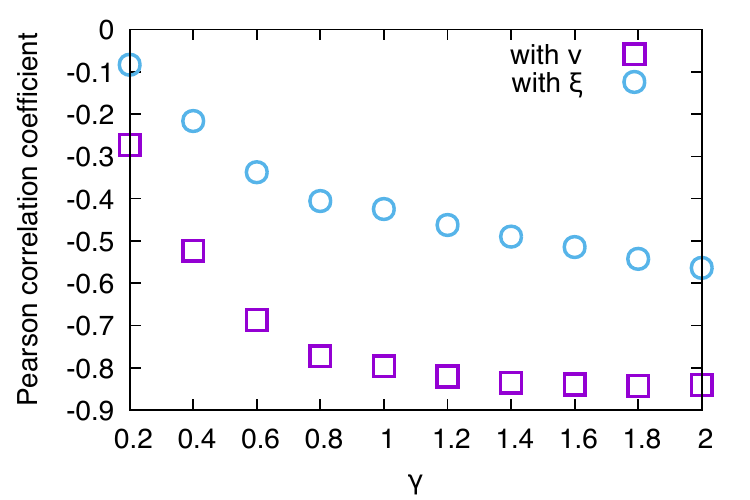}
\caption{The Pearson correlation coefficient between the modularity $Q$ in Eq.~\eqref{eq:modularity} and the nestedness measure $\nu$ in Eq.~\eqref{eq:NODF}, and that between the modularity $Q$ in Eq.~\eqref{eq:modularity} and the coreness measure $\xi$ in Eq.~\eqref{eq:NCQ_formula}.}
\label{fig:Pearson_corr_with_Q}
\end{figure}

그림~\ref{fig:Pearson_corr_with_Q}은 $89$개 상호 도움 네트워크에 대해 계산된 다양한 척도 $\gamma$에서의 군집도 (군집을 찾는 방법으로는 GenLouvain\cite{GenLouvain}을 이용하였다)와 포개짐 정도인 식~\eqref{eq:NODF}에서 정의한 $\nu$, 군집도와 중심-주변부 구분 정도인 식~\eqref{eq:NCQ_formula}에서 정의한 $\xi$ 사이의 피어슨 상관관계 지수 (Pearson correlation coefficient)를 나타낸 것이다. 보는 바와 같이 상관관계가 두 경우 모두 해상도 매개변수 $\gamma$가 증가할수록 단조 감소하고 있음을 알 수 있다. 일반적으로 특정한 척도에서 유의미한 군집 구조가 있다면 다른 중간 크기 성질인 포개짐 정도나 중심-주변부 구조와 관련이 있는 척도에서 상관관계가 두드러질 수 있는데, 그러한 경향이 없다는 뜻이다. 따라서 \ref{sec:MRF} 절에서 논의한 바와 같이 이것 또한 군집 구조가 이 네트워크들에서 큰 의미가 없을 수 있겠다는 또 하나의 단서가 된다.

\subsection{다양한 척도에서의 군집 구조 불일치도 분석}
\label{sec:PaI}

\begin{figure*}[t]
\begin{tabular}{ll}
(a) & (b) \\
\includegraphics[width=0.5\textwidth]{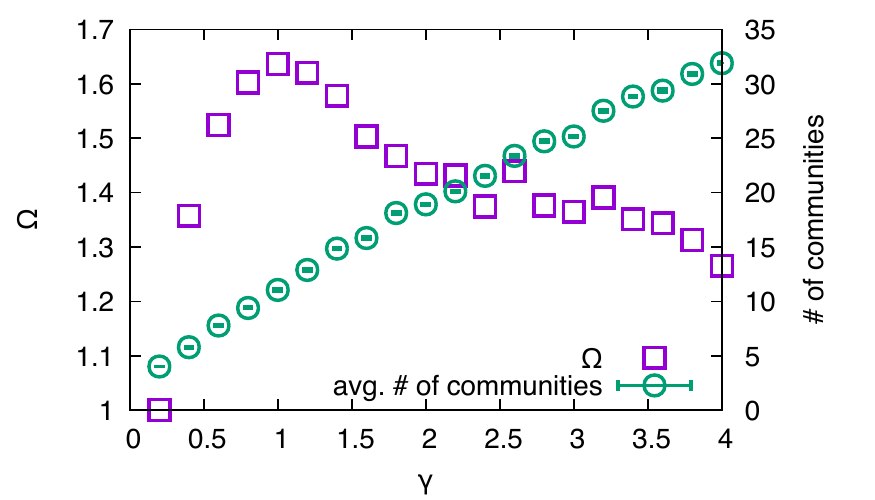} & \includegraphics[width=0.5\textwidth]{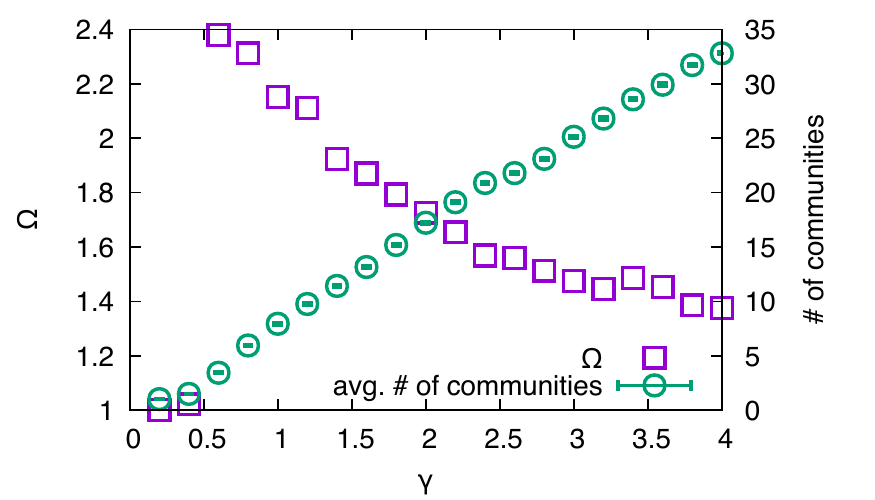} \\
(c) & (d) \\
\includegraphics[width=0.5\textwidth]{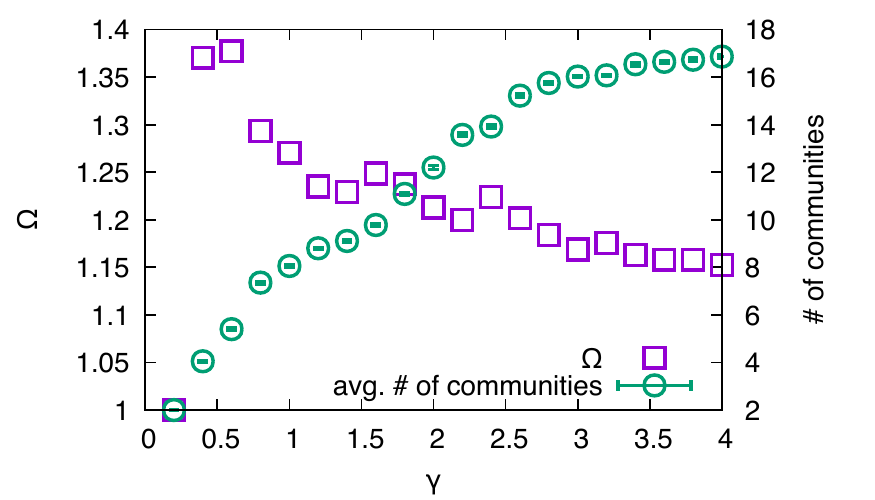} & \includegraphics[width=0.5\textwidth]{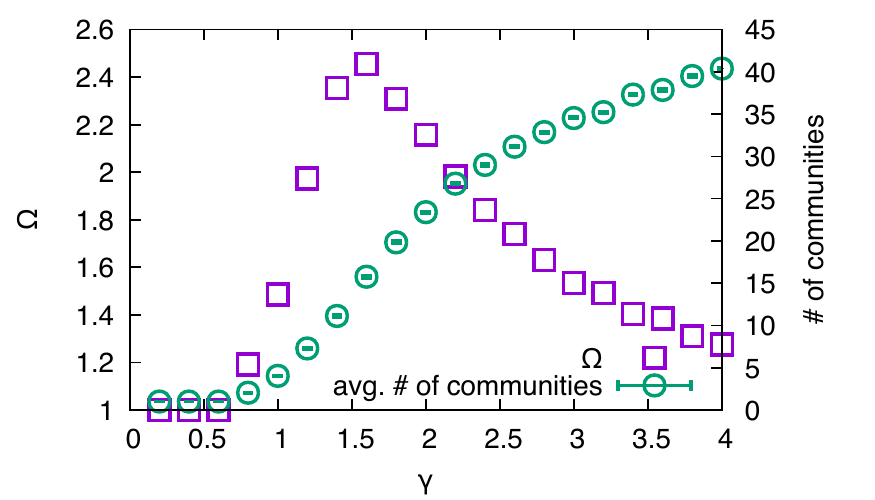} \\
\end{tabular}
\caption{The partition inconsistency $\Omega$ and the average number of communities found as the functions of the resolution parameter $\gamma$ in Eq.~\eqref{eq:modularity}, for (a) M\_PL\_001, (b) M\_PL\_018, (c) M\_PL\_027, and (d) M\_SD\_016.}
\label{fig:PaI_vs_gamma}
\end{figure*}

\begin{figure}[t]
\includegraphics[width=0.5\textwidth]{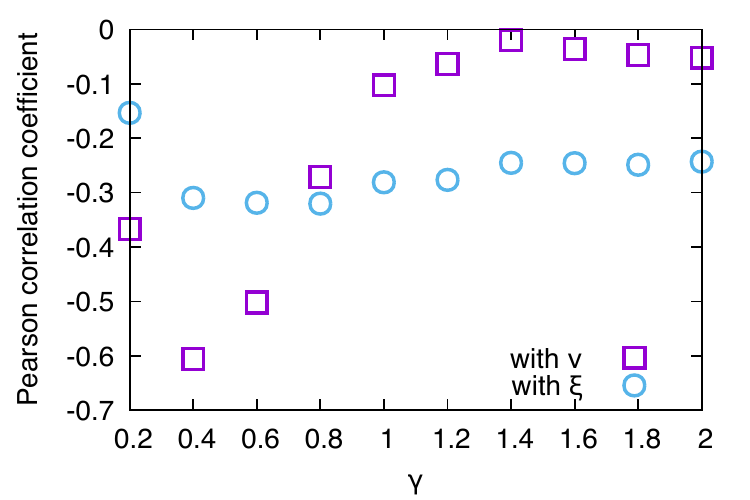}
\caption{The Pearson correlation coefficient between the partition inconsistency $\Omega$ in Eq.~\eqref{eq:PaI} and the nestedness measure $\nu$ in Eq.~\eqref{eq:NODF}, and that between the partition inconsistency $\Omega$ in Eq.~\eqref{eq:PaI} and the coreness measure $\xi$ in Eq.~\eqref{eq:NCQ_formula}.}
\label{fig:Pearson_corr_with_PaI}
\end{figure}

군집 구조 분석의 최종 단계로서, GenLouvain\cite{GenLouvain} 알고리즘의 확률적인 성질을 이용하여 다양한 척도에서의 군집 구조들 사이의 불일치도\cite{HKim2019,DLee2020}를 이용한 분석을 진행하였다. 우선, 주어진 해상도 매개변수 $\gamma$ 값에서 $m_\mathrm{tot}$번의 군집 구조 찾기를 했을 때 $m_a$번 출현하는 군집 구조 $a$, $m_b$번 출현하는 군집 구조 $b$ 사이의 성분 유사도 (element-centric similarity) $S_{ab}$를 참고문헌\cite{Gates2019}에 나온 공식으로 계산한 다음 각 군집 구조 $a$의 출현빈도 $p_a = m_a / m_\mathrm{tot}$를 이용하여 다음과 같이 전체적인 분할 불일치도 (partition inconsistency: PaI) $\Omega$를 계산하였다.
\begin{equation}
\Omega = \left( \sum_{a=1}^{m_C} \sum_{b=1}^{m_C} p_a p_b S_{ab} \right)^{-1} \,.
\label{eq:PaI}
\end{equation}
여기에서 $m_C$는 찾은 군집 구조들 중 서로 구분되는 다른 군집 구조의 개수이다. 식~\eqref{eq:PaI}에 따라 분할 불일치도는 찾아낸 모든 군집 구조들이 동일할 경우 최솟값인 $1$이 되며, 모든 군집 구조들이 서로 독립적이고 같은 확률로 나타날 경우 $a \neq b$인 경우에 대해 $S_{ab} = 0$이고 $a = b$인 경우에만 $S_{ab} = 1$, $p_a = p_b = 1/m_C$가 되어 최댓값인 $\Omega = ( \sum_{a=1}^{m_C} 1/m_C^2 )^{-1} = m_C$가 된다. 따라서 $\Omega$는 실질적으로 (effectively) 독립적인 군집 구조들의 개수로 해석할 수 있다. 즉 총 몇 가지 정도의 군집 구조들이 존재할 수 있는가를 추정한다고 볼 수 있으며, $\Omega$ 값이 작을수록 확률적인 군집찾기 알고리즘이지만 일관되게 (consistent) 군집을 찾는다는 뜻이므로 더 안정적인 또는 확실한 군집 구조라는 증거가 되는 것이다.

이러한 분할 불일치도가 다양한 척도에서 어떻게 나타나는지를, 각 척도에서 검출한 군집의 개수와 비교하면 어떤 척도에서 의미 있는 (안정적인) 군집 구조가 나타나는지를 파악할 수 있다\cite{DLee2020}. 그림~\ref{fig:PaI_vs_gamma}은 $89$개의 상호 도움 네트워크 중 대표적인 $4$가지 네트워크에서의 분할 불일치도 $\Omega$와 군집의 개수를 척도 $\gamma$에 따라 그린 것이다. 각 $\gamma$ 당 $100$번의 군집 구조 찾기를 시행하여 계산하였다. 공간의 제한 때문에 대표적인 $4$가지만 나타냈지만, 조사한 대부분의 네트워크들이 이러한 형태를 보이고 있다. 이러한 패턴은 특정 척도에 해당하는 군집들이 의미가 있을 경우에 나타나는, 군집 개수가 평평한 부분 (plateau)에서 분할 불일치도가 일시적으로 U자 형태로 낮아지는 (즉 그 U자의 중간 정도 부분에 해당되는 척도에서 일관적으로 확실한 군집 구조가 나타나는) 형태가 나오는 네트워크들\cite{DLee2020}과는 매우 다른 형태\footnote{그림~\ref{fig:PaI_vs_gamma}을 보면 분할 불일치도가 일시적으로 높아지는 부분은 관찰되는 경우들이 있는데, 일반적으로 의미있는 군집 구조가 있는 네트워크의 경우에는 높아졌다가 다시 낮아진 후 다시 높아지는 U 모양을 보여준다.}이며, 이 결과 역시 특정 척도에서 안정적으로 의미 있는 군집 구조를 찾지 못했다는 것을 뜻한다. 따라서, 앞서 소개한 결과들과 마찬가지로 이 생태계 상호 도움 네트워크들에서는 확실하게 의미 있는 군집 구조가 부재할 것이라는 추론에 힘을 실어주는 결과이다. 마지막으로, 분할 불일치도 $\Omega$와 포개짐 정도인 $\nu$, 분할 불일치도 $\Omega$와 중심-주변부 구분 정도인 $\xi$ 사이의 피어슨 상관관계 지수 (Pearson correlation coefficient)를 그림~\ref{fig:Pearson_corr_with_PaI}에 나타내었다. 이것으로부터도 역시 뚜렷한 상관관계 경향성을 볼 수 없음을 알 수 있다.

\section{결론}
\label{sec:conclusion}

``망치를 들고 있으면 모든 것이 못으로 보인다''\cite{BJKimBook}는 영어 속담처럼, 군집 구조가 가장 대표적인 네트워크의 중간 크기 성질이다보니 실제 세상 네트워크의 하부 구조를 분석할 때 연구자들이 대체로 가장 널리 알려진 성질인 군집 구조를 어떻게든 찾아보려는 경향이 있다. 생태계 네트워크도 예외가 아니어서, 포개진 구조와 군집 구조의 연관성을 주장하는 연구\cite{Fortuna2010}가 있었다. 적어도 본 연구에서 사용된 데이터에서는 특정 척도에서 의미 있는 안정적인 군집 구조를 찾기가 어려웠으며, 이것은 기본적으로 노드 개수가 많지 않은 소규모 네트워크들이라 생긴 통계적인 문제들일 수도 있다. 즉, 그 정도 규모의 생태계에서는 동등한 느낌의 집단이 나타나기 힘들며 그러한 생태계의 상호 도움 네트워크가 연결선수 분포, 중심-주변부, 일반-세부적 상호작용\footnote{이전 연구\cite{SHLee2016}에서도 밝혔듯이 그 셋은 모두 서로 깊은 관련이 있는 것으로 보인다.} 구분 등에서 모두 구조적으로 불평등한 구조로 되어 있다는 증거일 수도 있다. 

대부분의 실제 자연계, 사회계 네트워크 데이터의 응용으로서의 군집찾기 연구 결과들이 찾아낸 군집 구조의 특성과 의미에 대해서 논하는데, 본 연구를 통해 실제 그러한 군집들이 어느 정도 의미가 있는지를 먼저 살펴보는 것이 필요하지 않을까 하는 주의사항을 전달해 본다. 본 연구에서 사용된 것은 특정 데이터베이스로부터 나온 결론이기 때문에 크기가 훨씬 큰 척도에서 바라본 생태계 네트워크들에서는 군집 구조가 의미가 있을 수도 있다. 또한, 본 연구의 결론은 특정 척도에서의 군집 구조를 집어내기 (pinpoint)가 힘들다는 것이며 간접적인 정황 증거 (anecdotal evidence)의 성격이 짙기 때문에, 다양한 데이터와 다른 네트워크의 구조적 성질들을 추가 분석하는 후속 연구를 통해 좀 더 수학적이고 통계적인 결론 도출의 필요성이 있다는 말로 논문을 마무리한다.

\section*{감사의 글}
이 논문은 2020년도 경남과학기술대학교 교원 연구활성화 지원 사업의 예산지원으로 수행되었음. 분석에 필요한 코드들 중 일부를 제공한 Daniel J. Fenn 님 (\ref{sec:MRF} 절의 중간 크기 반응 함수 계산에 필요한 MATLAB 코드), Puck Rombach 님 (\ref{sec:correlation} 절의 중심-주변부 구조 계산에 필요한 MATLAB 코드), 김희태 님과 이대경 님 (\ref{sec:PaI} 절의 군집 구조 불일치도 계산에 필요한 python 코드 제공과 군집 구조 불일치도에 대한 논의)에게 감사한다.




 \end{document}